\documentclass[a4paper,twoside,twocolumn,english,prb,showpacs]{revtex4}
\usepackage[T1]{fontenc}
\usepackage[latin1]{inputenc}
\usepackage{graphicx}
\usepackage{amssymb}

\makeatletter

\providecommand{\tabularnewline}{\\}

\usepackage{babel}
\makeatother
\begin{document}

\title{Competition between local and nonlocal dissipation effects in two-dimensional
quantum Josephson junction arrays}

\author{T. P. Polak}

\address{Consiglio Nazionale delle Ricerche - Istituto Nazionale per la Fisica
della Materia, Complesso Universitario Monte S. Angelo, 80126 Naples,
Italy}

\address{Istituto di Cibernetica ''E.Caianiello'' del CNR, Via Campi Flegrei
34, I-80078 Pozzuoli, Italy.}

\email{polak@fisica.cib.na.cnr.it}

\author{T. K. Kope\'c}

\address{Institute for Low Temperatures and Structure Research, Polish Academy
of Sciences, POB 1410, 50-950 Wroclaw 2, Poland}

\email{kopec@int.pan.wroc.pl}


\begin{abstract}
We discuss the local and nonlocal dissipation effects on the existence
of the global phase coherence transitions in two dimensional Josephson-coupled
junctions. The quantum phase transitions are also examined for various
lattice geometries: square, triangular and honeycomb. The $T=0$ superconductor-insulator
phase transition is analyzed as a function of several control parameters
which include self-capacitance and junction capacitance and both local
and nonlocal dissipation effects. We found the critical value of the
nonlocal dissipation parameter $\alpha_{1}$ depends on a geometry
of the lattice. The critical value of the normal state conductance
seems to be difficult to obtain experimentally if we take into consideration
different damping mechanisms which are presented in real physical
systems.
\end{abstract}
\maketitle

\section{Introduction}

Macroscopic quantum effects in two-dimensional Josephson junction
arrays (JJA's) have been extensively studied both theoretically\cite{simanek1,doniach,wood,kopec1,jose,kopec2,ambegaokar}
and experimentally\cite{voss,wess,zant} during the last years. The
quantum nature of the phase of a superconducting order parameter is
reflected in phase transitions in JJA's. In nondissipative JJA's the
two main energy scales are set by the Josephson coupling $E_{J}$
between superconducting islands and the electrostatic energy $E_{C}$
arising from local deviations from charge neutrality. The ratio $E_{C}/E_{J}$
determines the relevance of the quantum fluctuations and when it increases
above a critical value, the phase order is destroyed and the array
turns into the insulator. For large capacitive coupling $E_{C}\gg E_{J}$
the system can be modeled by a renormalized classical two-dimensional
($2D$) $XY$ model. In the opposite limit, the energy cost for transferring
charges between neighboring islands in the array is so high that charges
tend to be localized. While the nature of the classical $2D$ $XY$
model is well understood, its quantum generalization still poses unsettled
issues.

Modern fabrication techniques allow one to make arrays of ultrasmall
superconducting islands separated by insulators. In such systems the
important factor which has a profound impact on the ground state of
the JJA's is dissipation caused by Ohmic resistors shunting the junctions\cite{chakravaraty,fisher,simanek3}
or quasiparticle tunneling through the junctions.\cite{eckern,kampf1,choi}
Despite several experiments with $2D$ JJA's\cite{takahide,rimberg}
and superconducting granular films\cite{yazdani} existence of the
dissipation driven transition and critical value of the normal state
conductance is at least questionable. 

Phase diagrams in quantum JJA's with both mechanisms of dissipation
Ohmic and quasiparticle were studied theoretically by Zaikin.\cite{zaikin}
Calculations done within the framework of the instanton technique
reveal a zero temperature phase diagram with two dissipative phase
transitions. The author claims there are regions on the phase diagram
where disordered phase and the classical Josephson effect could take
place. Cuccoli \emph{et al.}\cite{cuccoli} \emph{}presented an analytical
study based on the effective potential approach. They proposed a model
in which two different relaxation times lead to the conductance matrix
with resistive shunts to the ground and among islands. Despite of
these accurate analytical studies the problem of a theoretical explanation
of phase diagrams in JJA's in dissipative environment is still open.

Our previous theoretical work\cite{polak} in which the attention
was focused on local dissipation effects, predicted the existence
of the critical value of the dissipation parameter $\alpha=2$ independent
on geometry of a lattice and magnetic field. Other theoretical studies\cite{wagenblast,wagenblast 1,chakravaraty,fisher}
suggest a rather broad range of the critical values of the dissipation
parameter $\alpha=0.5,0.84,1,2$ which depends on the dimension of
the system and mechanism of the dissipation. It seems that an unambiguous
in experimental measurement of the critical value of the normal state
conductance is elusive. Several groups\cite{yagi,ootuka,rimberg,pentilla,takahide,yamaguchi,yamaguchi1}
using different experimental techniques obtained various critical
values of $\alpha=0.5,0.8,1$. To explain these theoretical and experimental
difficulties we propose a model in which local (caused by shunt resistors
connecting the islands to a ground) and nonlocal (shunt resistors
in parallel to the junctions) dissipation effects are considered.

The purpose of this paper is to investigate phase transitions at zero
temperature in two-dimensional capacitively coupled superconducting
arrays with emphasis on the competition of local $\alpha_{0}$ and
nonlocal $\alpha_{1}$ dissipation effects. The detailed phase boundary
crucially depends on the ratio of mutual to self-capacitances $C_{1}/C_{0}$
and specific planar geometry of the array.\cite{kopec1} Aware of
that fact we consider capacitive matrix $C_{ij}$ within the range
of parameters $C_{1}/C_{0}$ which can be adjusted to the experimental
samples. We analyze phase diagrams for three different lattices: square
($\square$), triangular ($\vartriangle$) and honeycomb (H). We want
to emphasize that our approximation cannot be used for analysis of
the Berezinski-Kosterlitz-Thouless transitions since it is appropriate
only for physical systems where long-range order appears. 

The outline of the rest of the paper is the following: In Sec. II
we define the model Hamiltonian, followed by its path integral formulation
in terms of the dimensionality dependent nonmean-field like approach.
In Sec III we present the zero-temperature phase diagram results for
different JJA's geometries. Finally, in Sec. IV we discuss our results
and their relevance to other theoretical and experimental works.

\section{Model}

We consider a two-dimensional Josephson junction array with lattice
sites $i$, characterized by superconducting phase $\phi_{i}$ in
dissipative environment. The corresponding Euclidean action reads:

\begin{equation}
\mathcal{S}=\mathcal{S}_{\mathrm{C}}+\mathcal{S}_{\mathrm{J}}+\mathcal{S}_{\mathrm{D}},\label{sum action}\end{equation}
where\begin{eqnarray}
\mathcal{S}_{\mathrm{C}} & = & \frac{1}{8e^{2}}\sum_{i,j}\int_{0}^{\beta}d\tau\left(\frac{\partial\phi_{i}}{\partial\tau}\right)C_{ij}\left(\frac{\partial\phi_{j}}{\partial\tau}\right),\nonumber \\
\mathcal{S}_{\mathrm{J}} & = & \sum_{\left\langle i,j\right\rangle }\int_{0}^{\beta}d\tau J_{ij}\left\{ 1-\cos\left[\phi_{i}\left(\tau\right)-\phi_{j}\left(\tau\right)\right]\right\} ,\nonumber \\
\mathcal{S}_{\mathrm{D}} & = & \frac{1}{2}\sum_{i,j}\int_{0}^{\beta}d\tau d\tau'\alpha_{ij}\left(\tau-\tau'\right)\left[\phi_{i}\left(\tau\right)-\phi_{j}\left(\tau'\right)\right]^{2}.\label{action}\end{eqnarray}
 and $\tau$ is the Matsubara's imaginary time ($0\leq\tau\leq1/k_{\mathrm{B}}T\equiv\beta$);
$T$ is temperature and $k_{\mathrm{B}}$ the Boltzmann constant ($\hbar=1$).
The first part of the action (\ref{action}) defines the electrostatic
energy where $C_{ij}$ is the capacitance matrix which is a geometric
property of the array. This matrix is usually approximated as a diagonal
(self-capacitance $C_{0}$) and a mutual one $C_{1}$ between nearest
neighbors. We can write a general expression for the $C_{ij}$ in
the following form:\begin{equation}
C_{ij}=\left\{ \begin{array}{c}
C_{0}+zC_{1}\quad\textrm{for}\quad i=j\\
-C_{1}\quad\textrm{for nearest neighbors}\end{array}\right.\label{capacitance matrix}\end{equation}
which holds for periodic structures in any dimension; $z$ is coordination
number of the network. The second term is the Josephson energy $E_{J}$
($J_{ij}\equiv E_{J}$ for $\left|i-j\right|=\left|d\right|$ and
zero otherwise). The vector $d$ forms a set of $z$ lattice translation
vectors, connecting a given site to its nearest neighbors. The Fourier
transformed wave-vector dependent Josephson couplings $J_{\mathbf{k}}$
are different for various lattices. The third part of the action $\mathcal{S}_{\mathrm{D}}$
describes the dissipation effects and $\alpha_{ij}\left(\tau-\tau'\right)$
is a dissipation matrix. We choose two independent damping mechanisms,
the on-site and the nearest-neighbor, because usually, the damping
is described in terms of shunt resistors $R_{0}$ connecting the islands
to a ground and shunt resistors in parallel to the junctions related
to $R_{1}$. We can write dissipation matrix similar to Eq. (\ref{capacitance matrix})
in a more closed form:\begin{equation}
\alpha_{ij}=\left(\alpha_{0}+z\alpha_{1}\right)\delta_{ij}-\alpha_{1}\sum_{d}\delta_{i,j+d}\label{dissipative matrix}\end{equation}
with the vector $d$ running over nearest neighboring islands. The
dimensionless parameters \begin{equation}
\alpha_{0}=\frac{R_{Q}}{R_{0}},\qquad\alpha_{1}=\frac{R_{Q}}{R_{1}}.\end{equation}
 describe strength of the local and nonlocal dissipation respectively,
where $R_{Q}=1/4e^{2}$ is quantum resistance.

\subsection{Method}

Most of existing analytical works on quantum JJA's have employed different
kinds of mean-field-like approximations which are not reliable for
treatment spatial and temporal quantum phase fluctuations. The model
in Eq. \ref{action} encodes the phase fluctuation algebra given by
Euclidean group $E_{2}$ defined by commutation relations between
particle $L_{i}$ and phase $P_{j}$ operators,\begin{eqnarray}
P_{j} & = & e^{i\phi_{j}},\nonumber \\
\left[L_{i},P_{j}\right] & = & -P_{i}\delta_{ij},\nonumber \\
\left[L_{i},P_{j}^{\dagger}\right] & = & P_{i}^{\dagger}\delta_{ij},\nonumber \\
\left[P_{i},P_{j}\right] & = & 0,\end{eqnarray}
with the conserved quantity (invariant of the $E_{2}$ algebra)\begin{equation}
P_{i}P_{i}^{\dagger}\equiv P_{xi}^{2}+P_{yi}^{2}=1.\label{spherical constraint 1}\end{equation}
 The proper theoretical treatment of the quantum JJA's must maintain
the constraint in Eq. \ref{spherical constraint 1}. A formulation
of the problem in terms of the spherical model initiated by Kope\'c
and Jos\'e\cite{kopec3} leads us to introduce the auxiliary complex
field $\psi_{i}$ which replaces the original operator $P_{i}$. Furthermore,
relaxing the original ''rigid'' constraint and imposing the weaker
spherical condition:

\begin{equation}
\sum_{i}P_{i}P_{i}^{\dagger}=N.\label{spherical constraint}\end{equation}
where $N$ is the number of lattice sites, allows us to implementation
the spherical constraint:

\begin{eqnarray}
\mathcal{Z} & = & \int\left[\mathcal{D}\psi\right]\delta\left(\sum_{i}\left|\psi_{i}\right|^{2}-N\right)e^{-\mathcal{S}_{\mathrm{J}}\left[\psi\right]}\nonumber \\
 &  & \times\int\left[\mathcal{D}\phi\right]e^{-\mathcal{S}_{\mathrm{C}+\mathrm{D}}\left[\phi\right]}\prod_{i}\delta\left[\mathrm{Re}\psi_{i}-P_{xi}\left(\phi\right)\right]\nonumber \\
 &  & \times\delta\left[\mathrm{Im}\psi_{i}-P_{yi}\left(\phi\right)\right].\end{eqnarray}
where $\left[\mathcal{D}\psi\right]=\prod_{i}\mathcal{D}\psi_{i}\mathcal{D}\psi_{j}^{*}$
and $\left[\mathcal{D}\phi\right]=\prod_{i}\mathcal{D}\phi_{i}$.
It is convenient to employ the functional Fourier representation of
the $\delta$ functional to enforce the spherical constraint in Eq.
(\ref{spherical constraint}): \begin{equation}
\delta\left[x\left(\tau\right)\right]=\int_{-i\infty}^{+i\infty}\left[\frac{\mathcal{D}\lambda}{2\pi i}\right]e^{\int_{0}^{\beta}d\tau\lambda\left(\tau\right)x\left(\tau\right)},\end{equation}
which introduces the Lagrange multiplier $\lambda\left(\tau\right)$
thus adding a quadratic term (in $\psi$ field) to the action in Eq.
(\ref{action}). The evaluation of the effective action in terms of
the $\psi$ to second order in $\psi_{i}$ gives the partition function
of the quantum spherical model (QSM)\begin{equation}
\mathcal{Z_{\mathrm{QSM}}}=\int\left[\mathcal{D}\psi\right]\delta\left(\sum_{i}\left|\psi_{i}\right|^{2}-N\right)e^{-\mathcal{S}\left[\psi\right]}\label{partition function}\end{equation}
where the effective action reads:\begin{eqnarray}
\mathcal{S}\mathrm{\left[\psi\right]} & = & \sum_{\left\langle i,j\right\rangle }\int_{0}^{\beta}d\tau d\tau'\left\{ \left[J_{ij}\left(\tau\right)\delta\left(\tau-\tau'\right)\right.\right.\nonumber \\
 & + & \left.\mathcal{W}_{ij}^{-1}\left(\tau,\tau'\right)-\lambda\left(\tau\right)\delta_{ij}\delta\left(\tau-\tau'\right)\right]\psi_{i}\psi_{j}^{*}\nonumber \\
 & + & \left.N\lambda\left(\tau\right)\delta\left(\tau-\tau'\right)\right\} .\label{sqsa}\end{eqnarray}
Furthermore, 

\begin{equation}
\mathcal{W}_{ij}\left(\tau,\tau'\right)=\frac{\delta_{ij}}{\mathcal{Z}_{0}}\int\left[\mathcal{D}\phi\right]e^{i\left[\phi_{i}\left(\tau\right)-\phi_{j}\left(\tau'\right)\right]}e^{-\mathcal{S}_{\mathrm{C}+\mathrm{D}}\left[\phi\right]},\label{correlator}\end{equation}
is the phase-phase correlation function with statistical sum

\begin{equation}
\mathcal{Z}_{0}=\int\left[\mathcal{D}\phi\right]e^{-\mathcal{S}_{\mathrm{C}+\mathrm{D}}\left[\phi\right]},\end{equation}
where action $\mathcal{S}_{\mathrm{C}+\mathrm{D}}\left[\phi\right]$
is just a sum of electrostatic and dissipative terms in Eq. (\ref{action}).
After introducing the Fourier transform of the field\begin{equation}
\phi_{i}\left(\tau\right)=\frac{1}{N\beta}\sum_{\mathbf{k}}\sum_{n=-\infty}^{+\infty}\phi_{\mathbf{k},n}e^{-i\left(\omega_{n}\tau-\mathbf{k}\mathbf{r}_{i}\right)}\end{equation}
with $\omega_{n}=2\pi n/\beta$, $\left(n=0,\pm1,\pm2,...\right)$
being the Bose Matsubara frequencies. From Eq. (\ref{correlator})
the phase-phase correlation function reads:

\begin{equation}
\mathcal{W}\left(\tau,\tau'\right)=\exp\left\{ -\frac{1}{\beta}\sum_{n\neq0}\frac{1-\cos\left[\omega_{n}\left(\tau-\tau'\right)\right]}{\frac{1}{8E_{C}}\omega_{n}^{2}+\frac{\alpha}{2\pi}\frac{J_{\mathbf{k}}}{E_{J}}\left|\omega_{n}\right|}\right\} .\label{correlation function}\end{equation}
 The charging energy parameter entering Eq. (\ref{correlation function})
is \begin{equation}
E_{C}=\frac{1}{2}e^{2}\left[\mathbf{C}^{-1}\right]_{ii}=\frac{e^{2}}{\pi\left(C_{0}+4C_{1}\right)}\mathbf{K}\left(\frac{4C_{1}}{C_{0}+4C_{1}}\right)\label{capacitance}\end{equation}
where $\mathbf{K}\left(x\right)$ is the elliptic integral of the
first kind\cite{abramovitz}. Furthermore we introduce quantities
$E_{0}=e^{2}/2C_{0}$ and $E_{1}=e^{2}/2C_{1}$ related to the island
and junction capacitances.

The dissipative parameter $\alpha$ and may be explicitly written
as\begin{equation}
\alpha^{-1}=\lim_{N\rightarrow\infty}\frac{1}{N}\sum_{\mathbf{k}}\frac{1}{\alpha_{0}+z\alpha_{1}-2\alpha_{1}\mathcal{E}_{\mathbf{k}}}\label{definition cap}\end{equation}
where $\mathcal{E}_{\mathbf{k}}$ is a dispersion and has different
form for various lattices. In the present paper we consider three
different geometries of the lattice: square ($\square$), triangular
($\vartriangle$) and honeycomb (H): \begin{eqnarray}
\mathcal{E}_{\mathbf{k}}^{\square} & = & \cos k_{x}+\cos k_{y},\nonumber \\
\mathcal{E}_{\mathbf{k}}^{\vartriangle} & = & \cos k_{x}+2\cos\left(\frac{k_{x}}{2}\right)\cos\left(\frac{\sqrt{3}}{2}k_{y}\right),\nonumber \\
\mathcal{E}_{\mathbf{k}}^{\mathrm{H}} & = & \sqrt{\frac{1}{2}\left(\frac{3}{2}+\mathcal{E}_{\mathbf{k}}^{\vartriangle}\right)}.\label{ek dependence}\end{eqnarray}
 with the lattice spacing set to $1$. The results of the sum over
wave vectors in Eq. \ref{definition cap} are placed in Appendix B.
Finally, for small frequencies, $\alpha_{0}\leq2$ and $\alpha_{1}\leq1$
the inverse of correlation function (\ref{correlation function})
becomes:\begin{equation}
\mathcal{W}^{-1}\left(\omega_{n}\right)=\left\{ \begin{array}{c}
\frac{1}{8E_{C}}\omega_{n}^{2}+\frac{\alpha}{2\pi}\frac{J_{\mathbf{k}}}{E_{J}}\left|\omega_{n}\right|\quad\textrm{for}\quad\omega_{n}\neq0\\
0\quad\textrm{otherwise}\end{array}\right.\end{equation}
In order to determine the Lagrange multiplayer $\lambda$ we observe
that in the thermodynamic limit ($N\rightarrow\infty$) the steepest
descent method becomes exact. The condition the integrand in Eq. (\ref{partition function})
has a saddle point $\lambda\left(\tau\right)=\lambda_{0}$, leads
to an implicit equation for $\lambda_{0}$:\begin{equation}
1=\frac{1}{\beta N}\sum_{\mathbf{k}}\sum_{n\neq0}G\left(\mathbf{k},\omega_{n}\right),\label{sph constraint}\end{equation}
where\begin{equation}
G^{-1}\left(\mathbf{k},\omega_{n}\right)=\lambda_{0}-J_{\mathbf{k}}+\frac{1}{8E_{C}}\omega_{n}^{2}+\frac{\alpha}{2\pi}\frac{J_{\mathbf{k}}}{E_{J}}\left|\omega_{n}\right|.\label{lagrange coefficient}\end{equation}
The emergence of the critical point in the model is signaled by the
condition\begin{equation}
G^{-1}\left(\mathbf{k}=0,\omega_{n}=0\right)\equiv\lambda_{0}-J_{0}=0\end{equation}
 which fixes the saddle point of the Lagrange multiplier $\lambda_{0}$
within the ordered phase $\lambda_{0}=J_{0}$.

\section{Phase diagrams}

\begin{figure}
\includegraphics[%
  scale=0.43]{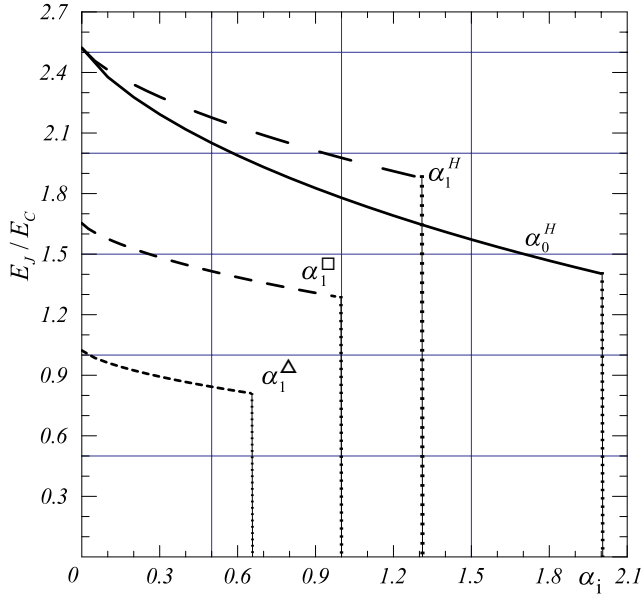}

\caption{Zero-temperature phase diagram for the total charging energy $E_{J}/E_{C}$
vs parameter of dissipation $\alpha_{i}$ ($i=1$ if $\alpha_{0}=0$
and $i=0$ if $\alpha_{1}=0$) for triangular ($\vartriangle$; $\alpha_{1}^{\mathrm{crit}}=2/3$),
square ($\square$; $\alpha_{1}^{\mathrm{crit}}=1$) and honeycomb
($H$; $\alpha_{1}^{\mathrm{crit}}=4/3$) lattice. Insulating (superconducting)
state is below (above) the curves. \label{0Tph}}
\end{figure}

A Fourier transform of the Green function in Eq. (\ref{lagrange coefficient})
enables one to write the spherical constraint (\ref{sph constraint})
explicitly as:

\begin{equation}
1=\frac{1}{\beta}\int_{-\infty}^{+\infty}d\xi\sum_{n\neq0}\frac{\rho\left(\xi\right)}{\lambda-\xi E_{J}+\frac{1}{8E_{C}}\omega_{n}^{2}+\frac{\alpha}{2\pi}\xi\left|\omega_{n}\right|}.\label{spherical constraint 2}\end{equation}
where\begin{equation}
\rho\left(\xi\right)=\frac{1}{N}\sum_{\mathbf{k}}\delta\left[\xi-\frac{J_{\mathbf{k}}}{E_{J}}\right]\end{equation}
is the density of states. We can easily see that solution of the model
requires the knowledge of the DOS for a specific lattice with the
superimposition of the self-consistency condition for the critical
line in Eq. (\ref{spherical constraint 2}). A Josephson-junction
array network is characterized for different lattices by the nearest-neighbor
Josephson coupling $E_{J}$ with the following wave-vector dependence\begin{eqnarray}
J_{\mathbf{k}}^{\square} & = & E_{J}\mathcal{E}_{\mathbf{k}}^{\square}\nonumber \\
J_{\mathbf{k}}^{\vartriangle} & = & E_{J}\mathcal{E}_{\mathbf{k}}^{\vartriangle}\nonumber \\
J_{\mathbf{k}}^{\mathrm{H}} & = & E_{J}\mathcal{E}_{\mathbf{k}}^{\mathrm{H}}\end{eqnarray}
 where $\mathcal{E}_{\mathbf{k}}$'s are given by Eq. \ref{ek dependence}.
The Fourier transform of the capacitance (dissipative) matrices for
a triangular and honeycomb lattice\cite{kopec1} can be also found
in Appendix C. 

By substituting the value of $\lambda_{0}=J_{\mathrm{max}}$ where
$J_{\mathrm{max}}$ denotes maximum value of the spectrum $J_{\mathbf{k}}$,
and after performing the summation over Matsubara frequencies, in
$T\rightarrow0$ limit we obtain the following result:

\begin{eqnarray}
1 & = & \frac{1}{\pi}\int_{-\infty}^{+\infty}d\xi\frac{\rho\left(\xi\right)}{\sqrt{\left(\frac{\alpha}{2\pi}\xi\right)^{2}-\frac{J_{\mathrm{max}}-\xi E_{J}}{2E_{C}}}}\nonumber \\
 &  & \times\ln\left[\frac{\frac{\alpha}{2\pi}\xi+\sqrt{\left(\frac{\alpha}{2\pi}\xi\right)^{2}-\frac{J_{\mathrm{max}}-\xi E_{J}}{2E_{C}}}}{\frac{\alpha}{2\pi}\xi-\sqrt{\left(\frac{\alpha}{2\pi}\xi\right)^{2}-\frac{J_{\mathrm{max}}-\xi E_{J}}{2E_{C}}}}\right].\label{critical line}\end{eqnarray}

\begin{figure}
\includegraphics[%
  scale=0.43]{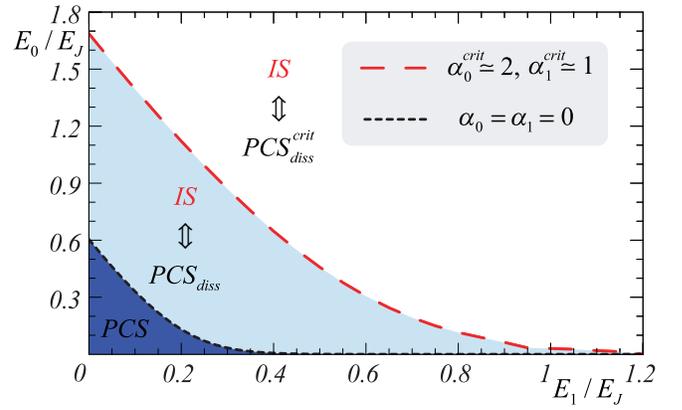}

\caption{Zero-temperature phase diagram for square 2D JJA's with self $C_{0}=e^{2}/2E_{0}$
and mutual $C_{1}=e^{2}/2E_{1}$ capacitance (Eq. \ref{capacitance})
for two values of local and nonlocal dissipation parameter $\alpha_{0}=\alpha_{1}=0$
and $\alpha_{0}=2$ and $\alpha_{1}=1$ (see Appendix). We can distinguish
three areas: phase coherent state ($PCS$) where phases in the islands
are well defined. Insulating state ($IS$) which could be driven to
the phase coherent state by effects of the dissipation ($PCS_{diss}$).
Finally, insulating state, where superconducting phase is perturbed
by strong zero point quantum fluctuations due to Coulomb blockade
that localizes charge carries to the islands. However system can be
driven to the phase coherent state ($PCS_{diss}^{crit}$) but only
by critical values of the dissipation parameters ($\alpha_{0}^{\mathrm{crit}}\simeq2$
and $\alpha_{1}^{\mathrm{crit}}\simeq1$).\label{Zero-temperature-phase 1}}
\end{figure}

\begin{figure}
\includegraphics[%
  scale=0.45]{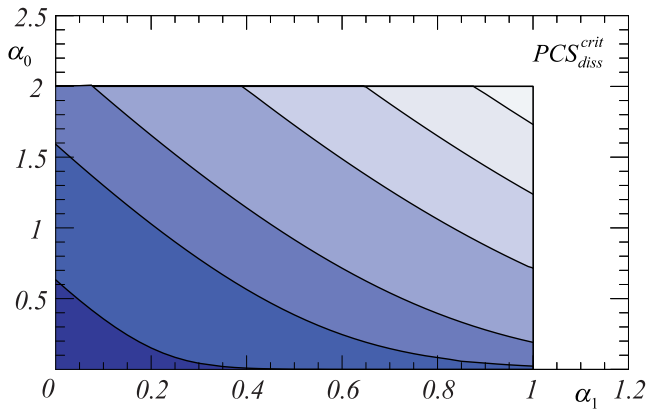}

\caption{Zero-temperature phase diagram for square 2D JJA's in space of local
$\alpha_{0}$ and nonlocal $\alpha_{1}$ dissipation parameters for
several values of the ratio $C_{1}/C_{0}$. From the top $C_{1}/C_{0}=1.25,1.67,2.5,5,10,50$.
Insulating (superconducting) state below (above) the curves.\label{Zero-temperature-phase 2}}
\end{figure}

The critical values of the nonlocal dissipation parameters have a
source in low temperature properties of the JJA's correlation function
in dissipative environment (Appendix A). The dependence of the critical
value $\alpha_{1}$ depicted in Fig. \ref{0Tph} is a direct result
of the divergence this phase-phase correlator. The Fig. \ref{Zero-temperature-phase 1}
and Fig. \ref{Zero-temperature-phase 2} point out the big difference
in values of the self $C_{0}$ and mutual $C_{1}$ capacitance and
competition between various dissipation mechanisms have a severe impact
on phase diagrams. In typical real situations mutual capacitance can
be at least two orders of magnitude larger than the self-capacitance
what indicates the samples are placed very close to $E_{1}/E_{J}$
axis in Fig. \ref{Zero-temperature-phase 1}. 

JJA's devoid of dissipation effects can be in two phases: insulator
phase ($IS$) and phase coherent state ($PCS$). However coupling
system to the environment we are able to drive arrays into $PCS$
even if localization of the charge carriers due to Coulomb blockade
is strong and dominates properties of the system. Furthermore for
each geometry of the lattice there are critical values of the dissipation
parameters that lead arrays to situation (Fig. \ref{Zero-temperature-phase 1})
where phases in the islands are well defined and quantum fluctuations
do not perturb a superconducting phase - region describe as $PCS_{diss}^{crit}$.
Notwithstanding between these two boundary situations there is a region
on the phase diagram in Fig. \ref{Zero-temperature-phase 1} where
concrete situation depends on the values of the parameters. In this
area $PCS_{diss}$ system can be driven to the $PCS$ but coupling
to the environment does not have to be so strong as in $PCS_{diss}^{crit}$
case. %

\section{Results}

Until now only three papers considered effects with both mechanisms
of the dissipation.\cite{zaikin,cuccoli,chakravaraty} The most interesting
is Cuccoli's work where authors introduced the full conductance matrix
for triangular and square lattices. It seems their results improve
the quantitative accuracy; nevertheless problem of the theoretical
explanation of the phase diagram of JJA's in dissipative environments
is thus open.

A model of an ordered array of resistively shunted Josephson junctions
was also considered by Chakravarty \emph{et. al.\cite{chakravaraty}}
and simplified at several points. They assumed that capacitance matrix
is diagonal $C_{ij}=C\delta_{ij}$. The authors claim the results
do not depend sensitively on detailed form of $C_{ij}$. Moreover
the matrix $\alpha_{ij}=h/4e^{2}R_{ij}$ where $R_{ij}$ is the shunting
resistance between grains $i$ and $j$ is reduced to the form in
which the information about the geometry of the lattice is not included.
The obtained zero-temperature phase diagram reveals the fact that
the critical value of the dissipation exists and is proportional to
the inverse of the dimension of the system which gives us critical
value $\alpha=1/2$ for a square lattice, but especially at low temperatures,
variational methods are not precise enough to perceive such a subtle
transition.

In our model the critical value of the nonlocal dissipation parameter
$\alpha_{1}$ behaves similarly. It depends on the maximum value of
the $J_{\mathbf{k}}$ spectrum. Because $J_{\mathbf{k}}$ exhibits
different characters for various lattices hence we could observe phenomenon
such as nonmonotonic dependence of the critical value of the nonlocal
dissipation parameter for various geometries of the array (see Fig.
\ref{0Tph}). When we assume diagonal form $\alpha_{ij}=\alpha_{0}\delta_{ij}$
then obviously our results will not change when we change the geometry
of the lattice simply because the shunt resistors connecting the islands
to ground are the same for each island. On the other hand if we take
into consideration that $\alpha_{0}=0$ and only $\alpha_{1}$ is
present, the situation changes because now values of the dissipative
matrix strongly depend on the $J_{\mathbf{k}}$ spectrum which indicates
various values of the matrix $\alpha_{ij}$ depend on the geometry
of the structure. This case is present in arrays with shunt resistors
in parallel to the junctions.

A standard way to study models with only diagonal charging energies
$C_{ij}=C\delta_{ij}$ corresponds to a complete absence of screening
by the other islands in the array.\cite{fisher1} Notice the mutual
capacitance can be at least two orders of magnitude larger than the
self-capacitance\cite{wess} $C_{1}\simeq10^{2}C_{0}$. We propose
a more realistic model in which both self and mutual capacitances
are nonzero. To see how significant this consideration is we shall
analyze Fig. \ref{Zero-temperature-phase 2}. If we assume, that $C_{1}/C_{0}\simeq50$
(see the lowermost line in Fig. \ref{Zero-temperature-phase 2}) we
can see the critical values of the dissipation parameters change dramatically
from local $\alpha_{0}=2$ and nonlocal $\alpha_{1}=1$ (Appendix
A) obtained for fewer realistic cases $C_{1}/C_{0}=1$ in which both
of the capacitances are comparable. In the limit ($C_{1}\neq0,C_{0}=0$)
the lattice model is equivalent to the Coulomb gas model with critical
properties not fully understood at present. However we have to emphasize
that the range of the Coulomb matrix becomes infinite when $C_{0}$
is set equal to zero. The phase boundary in Fig. \ref{Zero-temperature-phase 2}
shift downward with increasing ratio $C_{1}/C_{0}$ because lower
value of the Coulomb interaction between nearest neighbors reduces
strong quantum phase fluctuations and in consequence we observe a
growth of long-range phase coherence.

Yagi \emph{et al.\cite{yagi}} experimented with the superconductor-insulator
transition in two-dimensional network of Josephson junctions in detail
by varying the junctions-area. It was observed the critical tunneling
resistance exhibited significant junction area dependence. The low-temperature
behavior of the total charging energy $E_{J}/E_{C}$ as a function
$R_{Q}/R_{n}$, where $R_{n}$ is the tunneling resistance exhibits
the same behavior as the curve obtained from our theory for square
lattices (see Figure \ref{0Tph}) in the absence of the local dissipation
effects $\alpha_{0}=0$. The observed phase boundary, which is bending
downward, and the critical value of the nonlocal dissipation parameter
$\alpha_{1}^{\mathrm{crit}}=1$ is in excellent accordance with our
results.

In order to investigate the effects of quantum fluctuations and dissipation
in JJA's, another group\cite{takahide} made two-dimensional arrays
of small junctions with various $E_{J}/E_{C}$ and resistors which
caused dissipation effects. The value of $E_{J}$ was controlled by
varying the tunnel resistance. Each island was connected to the neighboring
ones by a shunt resistor as well as the tunnel junction. The resistance
of the shunt resistors was tuned by varying their length. Ground states
of $2D$ Josephson array in $E_{J}/E_{C}$-$R_{Q}/R_{s}$ parameters
space reveal the same behavior as previous experimental results but
there is a difference in critical value of the $R_{Q}/R_{s}\simeq0.5$
which also differs from value $\alpha_{1}^{\mathrm{crit}}$ obtained
in this paper. These discrepancies between experiments can be explained
in the framework of our model by taking into account different value
of the ratio junction-to-self capacitances which reduces the critical
value of the $R_{Q}/R_{s}$ and considering that not only nonlocal
dissipation mechanism is present $\alpha_{0}\neq0$ we can obtain
$R_{Q}/R_{s}\simeq0.5$ value (see Figure \ref{Zero-temperature-phase 2}).
Real numbers strongly depend on the properties of the junctions used
in experiments.

\section{Summary}

We have calculated quantum phase diagrams of two-dimensional Josephson
junction arrays using the spherical model approximation. The calculations
were performed for systems using experimentally attainable geometries
for the arrays such us square, triangular and honeycomb. The ground
state of the Josephson coupled array with a triangular lattice appears
to be most stable against the Coulomb effects. This geometry is also
the case in which the global coherent state emerged when the value
of the nonlocal dissipation parameter $\alpha_{1}$ is the lowest.
In JJA's we can observe the phase coherence transition which is caused
by electrostatic and dissipative effects. The detailed phase diagrams
crucially depend on the ratio junction-to-self capacitances, $C_{1}/C_{0}$
and both dissipation mechanism have a big impact on phase boundaries.
The nondiagonal terms in capacitive and dissipative matrices can change
the phase diagrams of the system drastically . It is necessary to
take them into considerations when we have different sources of dissipation
such us shunt resistors connecting the islands to a ground and shunt
resistors in parallel to the junctions. The experimental observation
of an universal resistance threshold for the onset of the global coherent
state seems possible, but appears to be difficult.

\begin{acknowledgments}
One of the authors wants to thank Dr. Ettore Sarnelli for carefully
reading manuscript and fruitful discussions. This work was supported
by the TRN {}``DeQUACS'' and some parts of it were done in Max-Planck-Institut
f\"ur Physik komplexer Systeme, N\"othnitzer Straße 38, 01187 Dresden,
Germany.
\end{acknowledgments}
\appendix

\section{Some Properties of the correlator}

Assume that $\alpha_{0}=0$ we write expression for the phase-phase
correlation function (similar to equation used in a previous calculations\cite{polak}
but modified by dissipative matrix) in form:\begin{equation}
\mathcal{W}\left(\tau\right)=\exp\left\{ -\frac{1}{\beta}\sum_{n\neq0}\frac{1-\cos\left(\omega_{n}\tau\right)}{\frac{1}{8E_{C}}\omega_{n}^{2}+\frac{\alpha_{1}}{2\pi}\frac{J_{\mathbf{k}}}{E_{J}}\left|\omega_{n}\right|}\right\} .\label{correlation function A}\end{equation}
It is easy to see the sum over $\omega_{n}$ is symmetric when we
change $\omega_{n}\rightarrow-\omega_{n}$. The key to obtain the
solution is a calculation the sum or the integral under the exponent
in Eq. (\ref{correlation function A}). Because we are going to investigate
low-temperature properties of the correlation function we could write
$\frac{1}{\beta}\sum_{\omega_{n}}\rightarrow\frac{1}{2\pi}\int_{-\infty}^{+\infty}d\omega$.
In that case (getting rid of abs) for large value $\tau$ we write\begin{eqnarray}
\mathcal{W}\left(\tau\right) & = & \exp\left[-\frac{1}{\pi}\int_{0}^{+\infty}d\omega\frac{1-\cos\left(\tau\omega\right)}{\frac{1}{8E_{C}}\omega^{2}+\frac{\alpha_{1}}{2\pi}\frac{J_{\mathbf{k}}}{E_{J}}\omega}\right]\nonumber \\
 & \simeq & \exp\left(-\frac{2\gamma E_{J}}{\alpha_{1}J_{\mathbf{k}}}\right)\left(\frac{\alpha_{1}J_{\mathbf{k}}E_{C}}{4\pi E_{J}}\left|\tau\right|\right)^{2E_{J}/\alpha_{1}J_{\mathbf{k}}}\end{eqnarray}
where $\gamma=0.57721$ is the Euler-Mascheroni constant.%
\begin{table}
\begin{tabular}{cccc}
\hline 
DOS &
$\vartriangle$&
H&
$\square$\tabularnewline
\hline 
$J_{\mathrm{max}}/E_{J}$&
$3$&
$\frac{3}{2}$&
2\tabularnewline
\hline
\end{tabular}

\caption{Maximum values of the spectrum $J\left(k\right)$ for three different
geometries of the lattices: triangular ($\vartriangle$), honeycomb
(H) and square ($\square$) \label{Maximum-values-of}}
\end{table}
Finally, after Fourier transform we see that correlator $\mathcal{W}^{-1}\left(\omega_{m}\right)\sim\left|\omega_{m}\right|^{2E_{J}/\alpha_{1}J_{\mathrm{max}}-1}$
at zero temperature diverges for $\alpha_{1}\geq2E_{J}/J_{\mathrm{max}}$.
Quantity $J_{\mathrm{max}}/E_{J}$ means the maximum value of the
$J_{\mathbf{k}}$ which differs for various lattices (see Table \ref{Maximum-values-of}).

\section{Dissipation parameter for considered Lattices}

In this appendix we give the explicit formulas for the dissipation
parameter discussed in Sec. II and III.

\subsection{Square lattice}

\begin{equation}
\alpha_{\square}^{-1}=\frac{2}{\pi\left(\alpha_{0}+4\alpha_{1}\right)}\mathbf{K}\left(\frac{4\alpha_{1}}{\alpha_{0}+4\alpha_{1}}\right)\end{equation}
where \begin{equation}
\mathbf{K}\left(x\right)=\int_{0}^{\pi/2}\frac{d\phi}{\sqrt{1-x^{2}\sin^{2}\phi}},\end{equation}
 is the elliptic integral of the first kind\cite{abramovitz} and
the unit step function is defined by:\begin{equation}
\Theta\left(x\right)=\left\{ \begin{array}{ccc}
1 & \textrm{for} & x>0\\
0 & \textrm{for} & x\leq0\end{array}\right..\end{equation}
For small values of the $\alpha_{1}$ we can write dissipation parameter
for square lattice as:\begin{equation}
\alpha_{\square}=\alpha_{0}+3\alpha_{1}-\frac{5}{4}\frac{\alpha_{1}}{\alpha_{0}}+\frac{9}{4}\frac{\alpha_{1}^{3}}{\alpha_{0}^{2}}+\mathcal{O}\left(\alpha_{1}^{4}\right),\end{equation}
for large values values of the $\alpha_{1}$: \begin{equation}
\alpha_{\square}=\frac{4\pi\alpha_{1}}{\ln\left(\frac{64\alpha_{1}}{\alpha_{0}}\right)}+\frac{1}{4}\pi\alpha_{0}\left[3-\frac{2}{\ln\left(\frac{64\alpha_{1}}{\alpha_{0}}\right)}\right]+\mathcal{O}\left(\frac{1}{\alpha_{1}}\right).\end{equation}

\subsection{Triangular lattice}

\begin{equation}
\alpha_{\bigtriangleup}^{-1}=\frac{1}{\pi\sqrt{3}}\frac{g}{\alpha_{1}}\mathbf{K}\left(\kappa\right)\end{equation}
where\begin{equation}
g=\frac{8}{\left[\left(2t+3\right)^{1/2}-1\right]^{3/2}\left[\left(2t+3\right)^{1/2}+3\right]^{1/2}}\end{equation}
\begin{equation}
\kappa=\frac{4\left(2t+3\right)^{1/4}}{\left[\left(2t+3\right)^{1/2}-1\right]^{3/2}\left[\left(2t+3\right)^{1/2}+3\right]^{1/2}}\end{equation}
with $t=\left(\alpha_{0}+6\alpha_{1}\right)/2\alpha_{1}$.

\subsection{Honeycomb lattice}

\begin{equation}
\alpha_{\mathrm{H}}^{-1}=\frac{1}{\pi\sqrt{3}}\frac{g}{\alpha_{1}}\frac{\alpha_{0}+3\alpha_{1}}{\alpha_{1}}\mathbf{K}\left(\kappa\right)\end{equation}
where\begin{equation}
g=\frac{8}{\left(2t-1\right)^{3/2}\left(2t+3\right)^{1/2}}\end{equation}
\begin{equation}
\kappa=\frac{4^{1/4}\left(2t\right)^{1/2}}{\left(2t-1\right)^{3/2}\left(2t+3\right)^{1/2}}\end{equation}
with $t=\left(\alpha_{0}+3\alpha_{1}\right)/2\alpha_{1}$.

\section{DOS for considered lattices}

In this appendix we give the explicit formulas for the density of
states discussed in Sec. II and III.

\subsection{Square lattice}

\begin{equation}
\rho^{\square}\left(\xi\right)=\frac{1}{\pi^{2}}\mathbf{K}\left(\sqrt{1-\left(\frac{\xi}{2}\right)^{2}}\right)\Theta\left(1-\left|\frac{\xi}{2}\right|\right),\end{equation}

\subsection{Triangular lattice}

\begin{equation}
\rho^{\bigtriangleup}\left(\xi\right)=\frac{2}{\pi^{2}\sqrt{\kappa_{0}}}\mathbf{K}\left(\sqrt{\frac{\kappa_{1}}{\kappa_{0}}}\right)\left[\Theta\left(\xi+\frac{3}{2}\right)-\Theta\left(\xi-3\right)\right],\end{equation}
where\begin{eqnarray}
\kappa_{0} & = & \left(3+2\sqrt{3+2\xi}-\xi^{2}\right)\left[\Theta\left(\xi+\frac{3}{2}\right)-\Theta\left(\xi+1\right)\right]\nonumber \\
 & + & 4\sqrt{3+2\xi}\left[\Theta\left(\xi+1\right)-\Theta\left(\xi-3\right)\right],\end{eqnarray}
\begin{eqnarray}
\kappa_{1} & = & 4\sqrt{3+2\xi}\left[\Theta\left(\xi+\frac{3}{2}\right)-\Theta\left(\xi+1\right)\right]\nonumber \\
 & + & \left(3+2\sqrt{3+2\xi}-\xi^{2}\right)\left[\Theta\left(\xi+1\right)-\Theta\left(\xi-3\right)\right].\end{eqnarray}

\subsection{Honeycomb lattice}

\begin{equation}
\rho^{\mathcal{\mathrm{H}}}\left(\xi\right)=4\left|\xi\right|\rho^{\bigtriangleup}\left(3-4\xi^{2}\right).\end{equation}

\end{document}